# Astro2020 Science White Paper

# Are Supernovae the Dust Producer in the Early Universe?

**Thematic Areas:** Planetary Systems Star and Planet Formation

Formation and Evolution of Compact Objects ☒ Cosmology and Fundamental Physics

☒ Stars and Stellar Evolution Resolved Stellar Populations and their Environments

Galaxy Evolution ☒Multi-Messenger Astronomy and Astrophysics

## Principal Author:

Name: Jeonghee Rho
Institution: SETI Institute
Email: jrho@seti.org
Phone: 650−961−6633

## Co-authors: (names and institutions)

Danny Milisavljevic (Purdue University), Arkaprabha Sarangi (NASA/GSFC), Raffaella Margutti (Northwestern University), Ryan Chornock (Ohio University), Armin Rest (Space Telescope Science Institute), Melissa Graham (University of Washington), J. Craig Wheeler (University of Texas Austin), Darren DePoy, Lifan Wang, Jennifer Marshall (Texas A&M University), Grant Williams (MMT Observatory), Rachel Street (Las Cumbres Observatory), Warren Skidmore (TMT International Observatory), Yan Haojing (University of Missouri-Columbia), Joshua Bloom (University of California, Berkeley), Sumner Starrfield (Arizona State University), Chien-Hsiu Lee (NOAO), Philip S. Cowperthwaite (Carnegie Observatories), Guy S. Stringfellow (University of Colorado, Boulder), Deanne Coppejans, Giacomo Terreran (Northwestern University), Niharika Sravan (Purdue University), Thomas R. Geballe (Gemini Observatory), Aneurin Evans (Keele University, UK) and Howie Marion (University of Texas, Austin)

## Abstract (optional):

Whether supernovae are a significant source of dust has been a long-standing debate. The large quantities of dust observed in high-redshift galaxies raise a fundamental question as to the origin of dust in the Universe since stars cannot have evolved to the AGB dust-producing phase in high-redshift galaxies. In contrast, supernovae occur within several millions of years after the onset of star formation. This white paper will focus on dust formation in SN ejecta with US-Extremely Large Telescope (ELT) perspective during the era of JWST and LSST.

## I. Introduction

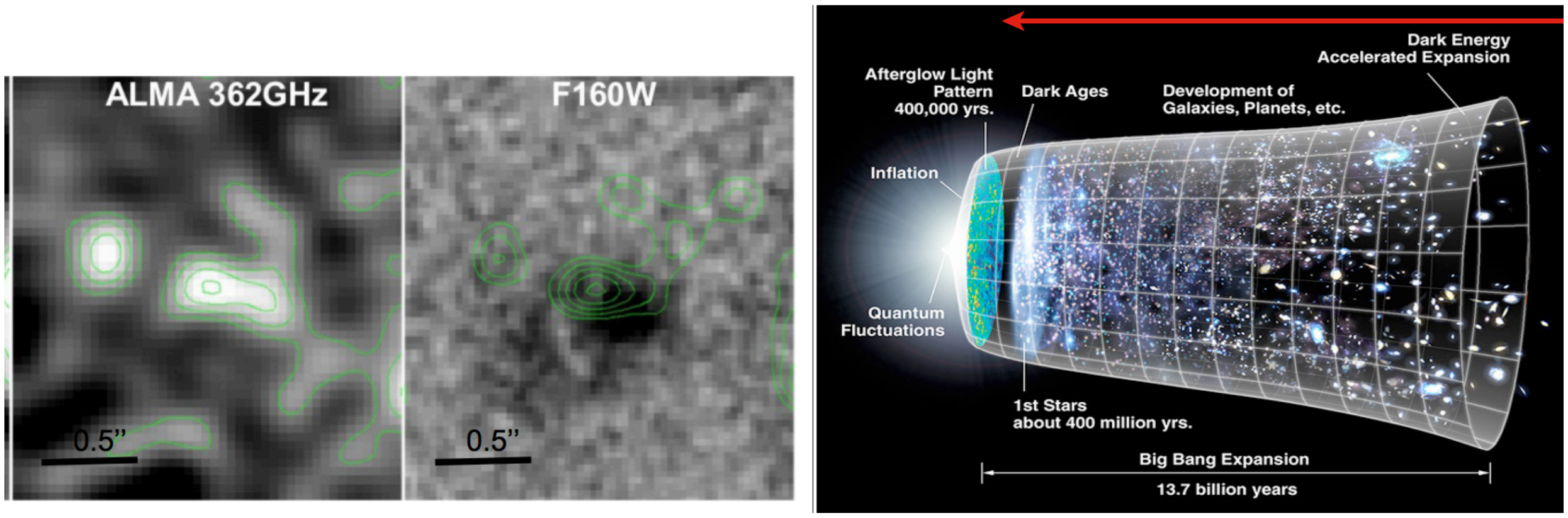

**Fig. 1.** (a: left) ALMA Band 7 continuum detection for the galaxy A2744_YD4 at z=8.34 [11]. (middle) HST F160W image with combined ALMA image contours overplotted [5]. (b: right) Diagram of the evolution of the early Universe (credit: WMAP). Are supernovae the dust producer in the early Universe after reionization occurs?

Whether or not supernovae (SNe) are a major source of dust has been a long-standing debate [2, 4, 6]. Deep submillimeter (submm) observations indicate that galaxies and QSOs have very large dust masses (>$10^8$ $M_\odot$) even at high redshifts of z > 5 [1, 11]. While we believe that most dust formation in current galaxies occurs in the stellar winds of asymptotic giant-branch (AGB) stars, it is difficult to explain the formation of such large amounts of dust in the early Universe (see Fig. 1b), since the timescales for stars to reach the AGB phase is longer than the age of the Universe being observed at these high redshifts [16, 23]. An ALMA detection of a gravitationally lensed galaxy at z~8 is shown in Fig. 1a.

Core-collapse supernova (ccSN) events, e.g. Type II, Ib and Ic, could explain the presence of dust in high-z galaxies, since their progenitors, stars much more massive than the sun, evolve on much shorter timescales (millions to a few tens of millions of years) and eject large amounts of heavy elements into the ISM. If a typical Type II SN were able to condense only 10% of its heavy elements into dust grains, it could eject around half a solar mass of dust into the ISM. The predicted dust mass formed in SNe depends on the progenitor mass; for a progenitor mass of 15 to 30 $M_\odot$, the predicted dust mass is from 0.1 to 1.0 $M_\odot$ [17, 23]. For a star formation rate 100 times that of the Milky Way this would lead to a dust mass of order $10^8 M_\odot$, if SNe and their remnants are not efficient dust destroyers [15, 17].

The temporal study of SNe is vital to understand the gap between the total starlight and the AGB starlight where ccSNe could dominate dust production in the Universe. *The amount of dust generated by Population III stars, which are thought to cause reionization, would significantly alter the star-formation process in early galaxies and stars.*

A number of studies of nearby SNe have been undertaken in order to estimate the mass of newly formed dust in the ejecta of ccSNe. Typical values derived for very young SNe (within ~ few years after the explosion) are only $10^{-4}$ – $10^{-2}$ $M_\odot$ of ejecta dust per SN event [10]. In contrast, infrared mapping of the young supernova remnant Cas A using *Herschel* and *Spitzer* confirmed that molecules and dust can form in ccSNe ejecta. Four young supernova remnants (SNRs), including SN 1987A, Cas A, the Crab Nebula and the SNR G54.1+0.3 [see Fig. 2a; 13, 3, 10, 8, 20] have dust masses of 0.1-0.9 $M_\odot$, in agreement with models. These results suggest SNe could be major dust factories at high-z galaxies.

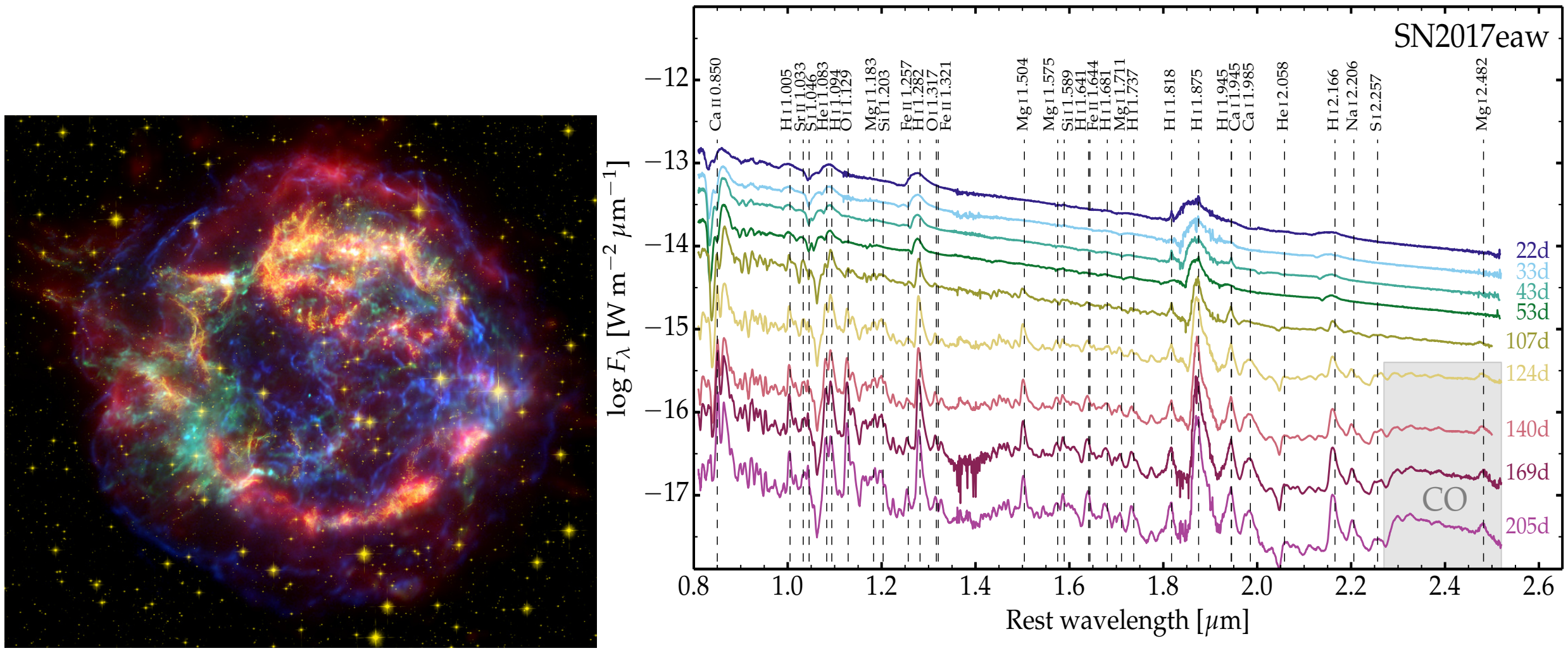

**Fig. 2.** (a: left) Mosaicked three-color image of young supernova remnant Cas A: *Spitzer* (red), *Hubble* (orange) and *Chandra* (blue). Infrared emission of Cas A is a representative case of dust formation [3]. (b: right) Gemini GNIRS near-IR 0.81 − 2.52 μm spectra (R~1700, dereddened) of SN 2017eaw obtained during 2017, in time order (top to bottom), offset and scaled differently for easier viewing. Wavelength interval with overtone CO emission is shaded [18].

The study of the role of ccSNe as major contributors to dust production in the Universe has grown dramatically over the last few years. However, a number of comprehensive studies of nearby SNe still find 1- 2 orders of magnitude smaller dust masses compared to the models and those found in YSNRs. While the evolution of CO at early times is remarkably consistent with chemically controlled dust models of Sarangi & Cherchneff [21], their dust model predicts an order of magnitude smaller amount of dust than previous models [17, 23, 24].

• What are the temporal evolutions of dust and CO formation and mass in ccSNe and how do they depend on SN progenitor type?

• Do molecules play a key role in dust formation in ccSNe ejecta? How does the physical state of the gas containing the CO change over time and how does it vary with progenitor mass?

• What types of grains (size, composition, etc.) form in SNe, and how do the dust mass and type evolve after the SN explosion? How are they related to the nucleosynthetic ejecta composition and progenitor mass?

• What fraction of the dust is destroyed from the passage of the reverse shock and how does that depend on the grain size? What are the implications to the ISM cycle in galaxies? Can SN knots of dense ejecta protect dust from the reverse shock with CO re-formation?

• Are SNe the source of dust in the early Universe? What is the efficiency of dust formation from

the ejecta mass? What implications are there for ccSN explosion models and nucleosynthesis from observations?

Supernovae (SNe) produce heavy metals, nuclear decay products, molecules and dust, which are the building blocks of planets and life. Meteoritic studies have shown that most presolar cosmic grains have condensed in the dense, warm stellar winds of evolved stars and in supernova explosions [14]. Dust that condensed deep within expanding supernova ejecta is believed to be responsible for some isotopic anomalies of heavy elements in meteorites. The study of supernovae to characterize dust composition and mass in terms of nucleosynthetic products will be an important guideline for searching for and characterizing high-z galaxies near the reionization of the Universe.

## 2. Near-IR studies of CO and dust formation in SNe

Carbon monoxide (CO) is one of the most powerful coolants in the ejecta of ccSNe and is believed to be responsible in large part for cooling the ejecta to temperatures at which dust can form. Dust production and type depends significantly on the C/O ratio in the gas, as well as the rate of destruction of CO, so measurements of CO are important tests of dust formation models. CO detection from SN1987A [12] in the near-infrared inferred onset of dust formation together with a drop in line intensities of refractory elements and the appearance of blue-shifted lines.

The easiest way to detect the onset of CO formation in SNe from the ground is via its first overtone at 2.3-2.4 μm. Detecting the fundamental band (4.5-5.0 μm) from the ground is more challenging, although that band is intrinsically considerably stronger. To date there are only about a dozen detections of CO (including Spitzer 4.5μm observations) in ccSNe [22]. Models of the evolution of CO and dust production exist [10, 21], but until last year the evolution of the CO and dust emission in ccSNe had only been followed in detail for SN1987A [12] 30 years ago! The evolution of the near-IR spectrum of the Type II-P SN 2017eaw in the nearby galaxy, NGC 6946 has been followed from Gemini N, obtaining eleven near-infrared (0.8 – 2.5 μm) spectra of the SN 2017eaw, spanning the time interval 22 – 494 days after discovery [see Fig. 2b; 18]. The first nine of these spectra (Fig. 2b) show the onset of CO formation at 2.3–2.4 μm and continuum emission at wavelengths greater than 2.1 μm from newly-formed hot dust. The observed CO masses are typically $10^{-4} – 10^{-3}$ $M_{\odot}$ during days 124 – 205. The timing of the appearance of CO and evolution of CO mass is remarkably consistent with chemically controlled dust models of Sarangi & Cherchneff. The near-infrared thermal emission of SN2010jl (see Fig. 3a) at later times (500 to 900 days) shows an accelerated growth in dust mass in the SN ejecta. The rapid growth of dust continuum in Fig. 3a demonstrates the powerful capability of studying dust formation in near-IR [7]. The wavelength-dependent extinction of the dust reveals the presence of very large (>1μm) grains, which resist destruction.

## 3. First Stars and Dust and CO Temporal Evolution of SNe with TMT+GMT

Future 30m class ELTs (such as TMT+GMT) will open to systematic investigation the currently uncharted territory of stellar explosions and dust formation in the ejecta in the early Universe potentially enabling the characterization of the explosions including *the first stars*. The sensitivity of the ELT program is shown in Figure 3b. The Pop III stars are believed to be massive (>100 $M_{\odot}$) [9]. The US ELTs is crucial for identifying the candidates of the first stars which can be followed-up at longer wavelength with JWST, while the JWST instruments a high sensitivity and high spatial resolution are available. A combination of US-ELT program and JWST will allow astronomy community to answer if SNe are the dust producer at the early Universe.

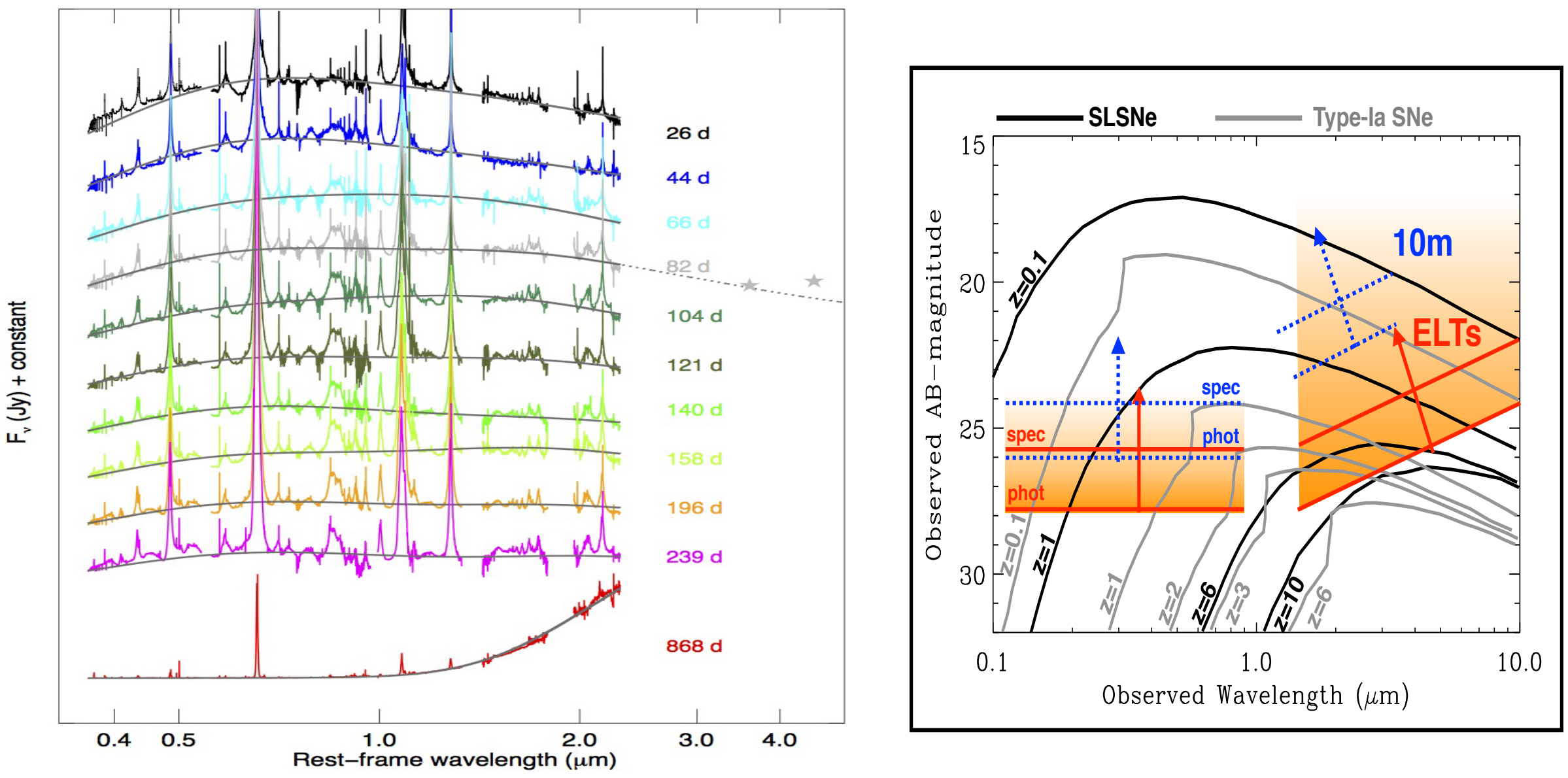

**Fig. 3.** (a: left) Near-IR spectra of SN2010jl demonstrate onset of dust formation can be detected in near-IR [7] (b: right) Optical-to-NIR SED of super luminous SNe and Type-Ia SNe at different distances. This KSP will open to systematic investigation the currently uncharted territory of stellar explosions in the early Universe potentially enabling the characterization of the explosions of the first stars (which might explode as super luminous (SL) SNe, black lines), or reveal the redshift evolution of our cosmic distance ladders (i.e. Type-Ia SNe, grey lines). Blue dotted lines: typical magnitude limits for a 10m aperture telescope for photometry (lower line) and spectroscopy (upper line). Red lines: same for the ELTs.

We expect the 2020 decade to be a golden age of transient domain astronomy (TDA). The Large Synoptic Survey Telescope (LSST) expects to discover 100,000 SNe per year which corresponds to ~100 transient objects per night. It is essential to have both TMT and GMT for their complementary instrument suites and ability to cover entire sky in the north and south as well as to maximize the benefits of LSST. ELT follow-up observations of the selected ccSNe will be crucial for revealing CO and dust formation and evolution of different types of SNe. The SNe that detect CO and dust evolution with ELTs will be the best targets for longer wavelength observation with JWST which will provide fine structure lines and continuum. *Timing is the key for TDA so that there is an urgent need for a US-ELT mission that reserves dedicated Target of Opportunity (ToO) time for TDA*.

ELTs enable to extend near-infrared observations of the formation and evolution of molecules and dust in newly exploded ccSNe. Only a dozen of SNe show evidence of dust and CO formation so far [6]. Bright, nearby (<10 Mpc) SNe have been observed in near-infrared to detect dust and CO. TMT+GMT from the near-IR high-resolution spectroscopic limit of ~26 magnitudes in K band will enable us to increase the sample of ccSNe. US-ELT program will be able to detect ccSNe up to the distance, 60 times further away from the current capability just for a short (15 min.) integration. By combining dramatic increases of newly discovered SNe with LSST, during the ELT era, we expect to be able to understand the onset of dust formation and its evolution as a function of time for different types of ccSNe.

Near-infrared instruments are critical to detecting CO molecules and dust continuum, as well as atomic lines, in particular, iron and cobalt lines at the later stage of SNe. IR spectrographs should be a priority for the US-ELT program, and an instrument with the capability of both near-IR and optical wavelengths is ideal for SN observations which help to understand nucleosynthesis and to detect isotopes (e.g. $^{12}C$ and $^{13}C$) together with dust formation in SNe. Broad spectroscopic observations follow-up generally require moderate resolutions (R~ 2000~4000) over the broadest possible wavelength range (0.32 – 2.4 μm). TMT/IRIS+NFIRAOS and GMT/GMTIFS or NIMOS+GMACS is potentially the best combination. The near-IR spectrometer with moderately high spectral resolution (R~ 2000-10,000) such as TMT/IRMOS and GMT/NIMOS or Near-IR echellette will be able to resolve the features of first overtone CO and the forest of nucleosynthetic element lines such as Fe, Co and Ni at nebular phase of SNe. We strongly advocate ARISE that has recently been proposed with a simultaneous optical/IR imager/spectrograph by the TMT Time Domain ISDT for the second generation suite of instrumentation for TMT, which would directly enable achieving the science goals outlined in this white paper. GMT/GMTIFS with a large field of view is ideal to observe ejecta knots in young SNRs.
Currently, only rare, brightest SNe and ones in nearby galaxies can be studied. We need the diversity of SN progenitor types, evolutionary stages and different environments such as different abundance and of SNe at higher redshift, and this requires larger aperture telescopes such as ELTs with IR spectrographs that can capture the CO bands and onset signatures of dust formation as well as numerous lines from heavy elements.

**Synergies with other missions:** The most abundant presolar grains and the predicted grains by dust models are SiC, amorphous silicates, forsterite ($Mg_2SiO_4$), enstatite ($MgSiO_3$), and corundum ($Al_2O_3$). JWST will likely reveal many of the dust features from these grains, which will help to identify dust composition. The uncertainty of dust composition is the highest uncertainty in estimating dust mass formed in SN ejecta. The far-IR space mission of Origins covers longer wavelengths and is important in determining accurate dust masses. The need for the far-IR capability of studying dust formation in SNe is addressed by another white paper (PI: Matsuura).